\documentclass[pra,twocolumn,showpacs,floatfix,superscriptaddress,aps]{revtex4}
\usepackage{amsmath}
\usepackage{makeidx}
\usepackage{amssymb}
\usepackage{graphicx}

\usepackage[usenames]{color}
\usepackage[normalem]{ulem}

\usepackage{hyperref}
\usepackage[T1]{fontenc} 
\usepackage{soul}
\begin{document}

\title{Fractional-charge vortex in a spinor Bose-Einstein condensate}

\author{Sandeep Gautam\footnote{sandeepgautam24@gmail.com}}
\author{S. K. Adhikari\footnote{adhikari44@yahoo.com, 
        URL  http://www.ift.unesp.br/users/adhikari}}
\affiliation{Instituto de F\'{\i}sica Te\'orica, Universidade Estadual
             Paulista - UNESP, \\ 01.140-070 S\~ao Paulo, S\~ao Paulo, Brazil}
      

\date{\today}
\begin{abstract} 
We classify all possible fractional charge vortices of  charge less than unity 
in spin-1 and spin-2 polar and cyclic Bose-Einstein condensates {(BECs)} with zero magnetization.  
Statics and dynamics of these vortices in  quasi-two-dimensional spinor BECs
are studied employing accurate numerical solution and a Lagrange variational 
approximation. The results for  density and collective-mode oscillation    
are illustrated using  fractional-charge BEC vortex of $^{23}$Na and {$^{87}$Rb} atoms with 
realistic interaction and trapping {potential} parameters.
\end{abstract}

\maketitle


\section{Introduction}

A scalar Bose-Einstein condensate (BEC) was first realized with alkali atoms of a single component 
of hyperfine spin $f$ \cite{Anderson}. In that case the spin degrees of freedom {do not} play any 
{essential} role in the dynamics. 
Later it has been possible to realize a spinor BEC of atoms of different spin components    
in an optical trap so that spin plays an essential role in the dynamics. For example, 
a spinor BEC can exhibit a spin-mixing dynamics, where atoms of one spin component may 
spontaneously change to atoms of another spin component. 
Depending on atomic {interactions} and the trap such a spin dynamics may 
spontaneously eliminate{,} from the spinor BEC{,} atoms of a certain spin component by 
transforming them to atoms of a different spin component.
In some cases, this transformation can go back and forth, thus leading to a 
{permanent} oscillation between atoms of different spin components \cite{ming}.  
Experimentally spinor BECs have
been realized with atoms of $^{23}$Na for $f=1$ \cite{Stenger}, $^{87}$Rb
for $f=1$ \cite{Chang} and $f=2$ \cite{Kuwamoto}, and
$^{52}$Cr for $f=3$ \cite{Pasquiou-1}. In the mean-field theory,
the $2f+1$ component wave function of spin-$f$ BEC is the solution of
a set of $2f+1$ coupled Gross-Pitaevskii (GP) equations \cite{Ohmi,Ho,Koashi,Ciobanu}.
In the absence of magnetic field, spin-1 and spin-2 BECs have, respectively,
two \cite{Ho} (ferromagnetic and antiferromagnetic or polar) and three \cite{Ciobanu} 
(ferromagnetic, antiferromagnetic or polar,
and cyclic) ground state phases.

Spinor BECs, unlike scalar BECs, can host a rich variety of topological excitations 
\cite{ueda, Stamper-Kurn,Masahito}. In scalar BECs, the single-valuedness of 
the wave function results in the emergence of vortices, under a complete 
rotation around which the phase (gauge phase) of the wave function
changes by an integer multiple $n$ of $2\pi$, {where $n$ equals the usual angular momentum quantum number.} 
The mass circulation for these vortices is quantized to 
 $nh/M$, where $M$ is the mass of an atom \cite{Onsager, Fetter}.
This is no longer the case for spinor BECs with non-zero spin-expectation per
particle, which is the case for the ferromagnetic phases of spin-1 and spin-2
BECs \cite{ueda,Masahito}. Here the mass circulation can change continuously by changing
spin configurations or spin texture as in the case of coreless vortices like
Mermin-Ho and Anderson-Toulouse 
vortices in spin-1 BEC \cite{Mizushima-1,Mizushima-2,Mizushima-3,Leanhardt}.
In the case of spin-1 polar \cite{Ho} and  spin-2 polar and cyclic \cite{Ciobanu}
 BECs, the change in the phase of the wave function
(gauge phase) after a complete rotation around the vortex core 
can be a fractional multiple $n$ of $2\pi$ resulting in the appearance of a fractional-charge vortex with mass circulation  equaling  a fractional multiple $n$ of $h/M$, where $n$ is the gauge  charge of the fractional vortex. One example of such a vortex is
 a half-quantum vortex or Alice vortex in a polar  spin-1 BEC
\cite{Zhou}. Besides these, both ferromagnetic and polar spin-1
BECs can host polar-core vortices {characterized by vortices with integer charges $\pm 1$ and 0 
in the spin components $m_f=\pm 1$ and 0, respectively, with the component $m_f=0$ lying in the vortex core created by components 
$m_f=\pm 1$}
\cite{Mizushima-2,Mizushima-3,Isoshima}. The spontaneous formation
of polar-core vortices in ferromagnetic spin-1 BEC of $^{87}$Rb has been experimentally
observed \cite{Sadler}. The generation of stable fractional-charge vortices in the
cyclic spin-2 BEC has also been theoretically investigated \cite{cyclic}.

In this paper, we study the statical and dynamical properties of a spin-1 polar BEC 
and {spin-2 polar and cyclic BECs}  with zero magnetization 
accommodating fractional-charge  vortices with gauge charge $n$ less than unity using a 
set of coupled mean-field GP equations.
In all cases, a quasi-two-dimensional (quasi-2D) \cite{Salasnich} trapping geometry will be employed.
Instead of studying a multi-component general state, we study how the 
ground states of these systems will dynamically evolve under rotation around $z$ axis. These ground states 
of lowest energy only populate two of the multi-component spin states and 
have a binary structure and are described 
by a set of binary mean-field GP equations.
For an analytic understanding, we present an approximation scheme based on 
a
Lagrange 
variational formulation with Gaussian {\it ansatz} for the component densities.  
The results of the variational approximation  are compared with an accurate  numerical solution of the mean-field model. 
Static properties of the axially symmetric vortex solutions of spin-2 BEC 
have been studied variationally by Pogosov {\it et al.} \cite{Pogosov}. However, that study did not consider the 
fractional-charge vortices.

 The paper 
is organized as follows. In Sec. \ref{sec-II}, we describe the coupled GP equations
for the spin-1 and spin-2 BECs. 
In Sec. \ref{3}, we present a classification of different fractional-charge vortices 
of spin-1 and spin-2 BECs which we study in this paper.
Here, the nonlinearities of the equivalent binary
system under rotation are defined in terms of the nonlinearities of the  
full $(2f+1)$-component GP equation. 
In Sec. \ref{sec-IV}, we present a dynamical Lagrange  variational 
analysis  to study the fractional-charge  spin-1 and spin-2 vortices using a 
set of binary GP equations where one or both components host a vortex. 
The numerical and variational results for the statics and dynamics of fractional-charge 
spin-1 and spin-2 BECs are reported in Sec. \ref{V}. 
The paper is concluded by a brief summary and discussion in Sec. \ref{VI}.

\section{GP equations for spin-1 and spin-2 BECs}
\label{sec-II}
When the trapping frequency along one axis, say $\omega_z$, is much larger 
than the geometric mean of other two, i.e., 
$\omega_z\gg\sqrt{\omega_x \omega_y}$, a spin-1 quasi-2D BEC of $N$ atoms of mass 
$M$ each can be described by the following set of coupled two-dimensional  
GP equations for different spin components $m_f=\pm 1,0$ \cite{ueda,Ho,spin1}
 \begin{eqnarray}&&
\mu_{\pm 1} \phi_{\pm 1}(\mathbf r) =
 {\cal H}\phi_{\pm 1}(\mathbf r) 
\pm   c^{}_1F_z\phi_{\pm 1}(\mathbf r) 
+ \frac{c^{}_1}{\sqrt{2}} F_{\mp}\phi_0(\mathbf r),
 \label{gps-1}\\
&&\mu_0 \phi_0(\mathbf r) =
{\cal H}\phi_0(\mathbf r)  
+ \frac{c_1}{\sqrt 2} [F_{-}\phi_{-1}(\mathbf r) 
+F_{+}\phi_{+1}(\mathbf r)]
\label{gps-2}, 
\end{eqnarray}
where ${\bf F}\equiv\{F_x,F_y,F_z\}$ is a vector with expectation
value of the three spin-operators over the multicomponent wavefunction as the three
components. It is termed as the spin-expectation value \cite{ueda}, and  
\begin{align}&
F_{\pm}\equiv  F_x \pm i F_y=
\sqrt 2[\phi_{\pm 1}^*(\mathbf r)\phi_0(\mathbf r)
+\phi_0^*(\mathbf r)\phi_{\mp 1}(\mathbf r)]\label{fpmspin1}, \\
&  F_z= \rho_{+1}(\mathbf r)-\rho_{-1}(\mathbf r)\label{fzspin1},
\quad
{\cal H}= -\frac{\nabla^2}{2} +{V}({\mathbf r})+c_0 \rho,\\
&c_0 = \frac{2N \sqrt{2\pi \gamma}(a_0+2 a_2)}{3 l_0},~
c_1 = \frac{2N \sqrt{2\pi \gamma}({a_2-a_0})}{3 l_0},\\
&\nabla^2=\frac{\partial ^2}{\partial x^2} + \frac{\partial ^2}{\partial y^2},~
V(\mathbf r)=\frac{x^2+\beta^2y^2}{2},~
{\mathbf r }\equiv\{x,y\}, \label{nabla_q2d}
\end{align}
where  the component density $
\rho_j=|\phi_j(\mathbf r)|^2$ with $j=\pm 1, 0$, the total density  $\rho=\sum_{j}\rho_j,$ and $\mu_{\pm 1},\mu_0$
are the respective chemical potentials and $^*$ denotes complex conjugate.
The component wavefunctions $\phi_j$'s satisfy the normalization condition $\int \sum_j \rho_j d{\bf r} =1$.
Here $a_0$ and $a_2$ are the $s$-wave scattering lengths in the 
total spin 0 and 2 channels,
 $l_0=\sqrt{\hbar/(M\omega_x)}$, $\beta = \omega_y/\omega_x$, 
$\gamma = \omega_z/\omega_x$, where $\omega_x,\omega_y,\omega_z$
are the confining trap frequencies in $x,y,z$ directions, respectively. 
Here length is measured in units of $l_0$, density in units of $l_0^{-2}$
and chemical potential in units of $\hbar\omega_x$.

 Similarly, the dimensionless coupled GP equations for different spin components
$m_f=\pm 2,\pm 1,0$, for a spin-2 
BEC can be written as \cite{Koashi,ueda,spin2}
 \begin{align} 
 \mu_{\pm 2}& \phi_{\pm 2}(\mathbf r) =
{\cal H}\phi_{\pm 2}(\mathbf r) 
+({{c}_2}/{\sqrt{5}}){\Theta}\phi_{\mp 2}^*(\mathbf r)\nonumber\\
&+{c}_1\big[{F}_{\mp} \phi_{\pm 1}(\mathbf r)\pm 2{F}_{{z}}\phi_{\pm 2}(\mathbf r)\big] 
\label{gp_s1},
\end{align}
\begin{align}
 \mu_{\pm 1}& \phi_{\pm 1}(\mathbf r) = 
{\cal H} \phi_{\pm 1} (\mathbf r) 
-({{c}_2}/{\sqrt{5}}){\Theta}\phi_{\mp 1}^*(\mathbf r)\nonumber\\
 &+{c}_1\big[\sqrt{3/2} {F}_{\mp}\phi_0(\mathbf r)+{F}_{\pm}\phi_{\pm 2} (\mathbf r)
\pm {F}_{{z}}\phi_{\pm 1}(\mathbf r)\big] ,
 \label{gp_s2}
 \\
 \mu_{0}& \phi_{0}(\mathbf r) =
{\cal H}\phi_0(\mathbf r) 
+({{c}_2}/{\sqrt{5}}){\Theta}\phi_{0}^*(\mathbf r)\nonumber\\
  &+c_1\sqrt{3/2}\big[{F}_{-}
 \phi_{-1}(\mathbf r)+{F}_{+}\phi_{+1}(\mathbf r)\big]
  ,\label{gp_s3}
\end{align}
where
\begin{align}
 {F}_{+} =&  {F}_{-}^*= 2(\phi_{+2}^*\phi_{+1}+\phi_{-1}^*\phi_{-2}) \nonumber \\
&+\sqrt{6}(\phi_{+1}^*\phi_0 +\phi_0^*\phi_{-1}),  \\
{F}_{{z}} =& 2(\rho_{+2}-\rho_{-2}) + \rho_{+1}-\rho_{-1},  \\
{\Theta} =& \frac{2\phi_{+2}\phi_{-2}-2\phi_{+1}\phi_{-1}+\phi_0^2}{\sqrt{5}}.\label{theta}
\end{align}    
Here $c_0 = 2 N \sqrt{2\pi\gamma}(4a_2+3a_4)/(7l_0), c_1=2 N\sqrt{2\pi\gamma}(a_4-a_2)/(7l_0),
c_2 = 2 N\sqrt{2\pi\gamma}(7a_0-10a_2+3a_4)/(7l_0)$, 
 $a_0,a_2$, and   $a_4$ are the $s$-wave scattering lengths in the 
total spin 0, 2 and 4 channels,
and $\mu_{\pm 2}, \mu_{\pm 1},$ $\mu_0$ are the respective chemical potentials. All repeated variables have 
the same meaning as  in the spin-1 case.

Depending on the values of the interaction parameters $c_j$, the spinor BEC acquires 
distinct properties and is classified as ferromagnetic, antiferromagnetic or polar, cyclic, etc.  
For a spin-1 BEC, $c_1 <0$ corresponds to ferromagnetic phase and $c_1>0$ to polar phase \cite{Ho}. For a spin-2 BEC,   $c_1<0$ and $c_2>20c_1$ correspond to ferromagnetic phase 
and $c_2<0$,  $c_2<20c_1$ correspond to polar phase, and $c_1>0$ and $c_2>0$ correspond to cyclic phase \cite{Ciobanu,ueda}.

\section{Classification of fractional-charge vortices}
\label{3}
 
 For a spinor  BEC with zero {spin-expectation value}, 
${\bf F} = 0$, the superfluid velocity 
  ${\bf v_s}$
is 
proportional to the gradient of the gauge phase $\theta$ of the wave function, i.e. ${\bf v_s} = \hbar \nabla \theta/M$. 
The same is true  in a scalar BEC. Hence the spinor BECs with ${\bf F} = 0$ 
(spin-1 and spin-2 polar and spin-2 cyclic BECs)
are irrotational ($\nabla \times \bf v_s = 0$) 
and the circulation  {of the velocity field} is quantized \cite{ueda}:  
 $\oint {\bf v_s}.d{\bf l} = n h/M $, where $n$ is an integer or a   rational fraction.
On the other hand, for spin-1 and spin-2 ferromagnetic BECs,  the circulation is
not quantized \cite{ueda}. The vortices in spin-1 and spin-2 polar and spin-2 cyclic BECs 
with the circulation of the velocity field equal to fractional multiple of $h/M$ are 
known as fractional-charge vortices, 
which emerge due to
the fact that a spinor BEC has SO(3) rotational symmetry in addition to U(1) global
gauge symmetry of a scalar BEC \cite{sakurai,ueda}.

For  a spinor,  rotation in spin space is generated by  
$e^{i\varphi}, \varphi \equiv -\alpha  S_z/\hbar,$ where $\alpha$ is the azimuthal angle of rotation about $z$ axis 
and  $S_z$ is the spin projection.  In this study of votices we consider 
rotation around $z$ direction only as this is the only relevant rotation in the present 
quasi-2D model confined to  the $x-y$ plane, where the dynamics along the $z$ direction is frozen.
Normal wave function and spinor must have the same value for 
 $\alpha =0$ and $2 \pi$.
A part of the single-valuedness  may come from 
the gauge phase ($e^{i\theta}$) and a part from the spin phase ($e^{i\varphi}$).  
 
\subsection{Spin-1/2 particle}
How  a fractional charge emerges is already explicit in the dynamics of a fundamental spin-1/2 particle 
under rotation using the Pauli formalism. A quantum mechanical spin-1/2 state 
is described by a complex-valued vector with two components called a spinor, 
which has a distinct behavior under rotation when compared with that of a 
spin-0 system.  In the Pauli formalism of the spin-1/2 system, the $2\times 2$ 
matrix representation of the rotation 
operation around direction $z$ can be written as \cite{sakurai}
 \begin{align}{\cal D}(\alpha)=&
e^{{-iS_z\alpha}/{\hbar}}\equiv e^{{-i\sigma_z\alpha}/{2}}\nonumber \\=&
\left( {\begin{array}{cc}
   1 & 0 \\       0 & 1 \      \end{array} } \right)\cos\left(\frac{\alpha}{2}\right)-i\left( {\begin{array}{cc}
   1 & 0 \\       0 & -1 \      \end{array} } \right)\sin\left(\frac{\alpha}{2}\right)
\nonumber \\ =& \left( {\begin{array}{cc}
   e^{-i\alpha/2} & 0 \\       0 &    e^{i\alpha/2} \      \end{array} } \right),
\end{align}
where   $\sigma_z\equiv 2S_z/\hbar$ is the  $2\times 2$ Pauli matrix. 
This rotation operator has the property 
${\cal D}(\alpha+2\pi)=-{\cal D}(\alpha), {\cal D}(\alpha+4\pi)={\cal D}(\alpha)$, 
which results in    peculiar properties of a spin-1/2 spinor. Under a rotation of $2\pi$,
 the spinor changes sign and remains unchanged under a rotation of $4\pi$.
A generic spinor is obtained as $\chi= {\cal D}(\alpha) \chi_0$, where $\chi_0=(1/\sqrt 2,1/\sqrt 2)^T$ is 
a fundamental state  with equal probability in the two components (zero magnetization).
It is customary to define a spinor  
with an extra Gauge phase of $e^{i\theta}$ to obtain    $e^{i\theta}(e^{-i\alpha/2}/\sqrt{2},e^{i\alpha/2}/\sqrt{2})  ^T$. To
 guarantee the single-valuedness under 
a rotation of $2\pi$, we take $\alpha =\beta$ and $\theta=\beta/2$ and obtain the spinor 
 $e^{i\beta/2}(e^{-i\beta/2}/\sqrt{2},e^{i\beta/2}/\sqrt{2})  ^T$.
As is evident,
the gauge phase can be scaled out of the wave function, whereas
the spin phases are inherent to the wave function and cannot be scaled out.
This form of Gauge phase corresponds to an angular momentum 1/2, and 
the single-valuedness of the spinor under a rotation of $2\pi$ demands  that the Gauge charge 
of the spinor has the fractional value 1/2. There are many more interesting possibilities of
fractional charge for spinor BECs of spin 1 and 2 as we see below.

\subsection{Spin-1 BEC}

{ There are two degenerate ground states of the spin-1 BEC with zero (longitudinal) magnetization
 ($ {\cal M} = \int F_z({\bf r}) d\mathbf r= 0$) \cite{tf}.
The first
of these states has $\phi_0 = 0$.
The rotation operator around direction $z$ in this case can be written as \cite{ueda}
\begin{align}
{\cal D}(\alpha)=\left( {\begin{array}{ccc}
   e^{-i\alpha} & 0&0 \\ 0&1&0 \\       0 & 0&   e^{i\alpha} \      \end{array} } \right)
\end{align}
 A general normalized state with zero magnetization and $\phi_0=0$ can be obtained 
by operating ${\cal D}(\alpha)$ on the state $\chi_0=(1/\sqrt 2,0,1/\sqrt 2)^T$ and is 
expressed
as $ e^{i\theta}(e^{-i\alpha}/\sqrt{2},0,e^{i\alpha}/\sqrt{2})^T$.  
The simplest way to maintain a single-valued spinor with a fractional gauge charge ($\theta \ne 0$)
is to take $\theta=\alpha= \beta/2$. The wave function, then,  becomes
$e^{i\beta/2}(e^{-i\beta/2}/\sqrt{2}, 0,e^{i\beta/2}/\sqrt{2})  ^T$. 
This state is very similar to the spin-1/2 spinor considered above. 
This is a 1/2-1/2 vortex meaning 1/2 unit of gauge charge and 1/2
unit of spin charge. Because of 1/2 unit of gauge charge, this vortex is classified as a vortex of fractional charge 1/2.
In numerical calculation, it will become a vortex of charge 0 in component
$\phi_{+1}$  and charge 1 in component $\phi_{-1}$. The second degenerate ground state
has $\phi_{\pm 1} = 0$ and can not host a vortex of fractional charge. Hence, 1/2-1/2 vortex
is the only possibility for  a fractional vortex of charge less than one for a
spin-1 BEC. The degeneracy between the two states is broken at a non-zero
value of  {magnetization $ {\cal M}  $ with the 
first state  ($\phi_0 = 0$) emerging as the ground state.}

\subsection{Spin-2 BEC}

There are a few different possibilities of polar and cyclic ground  states for a spin-2 BEC
with zero magnetization, which lead to different types of fractional-charge vortices, of which we describe below the ones with charge less than unity. 

(a) There are three degenerate ground states of a non-rotating 
spin-2 polar
BEC with zero magnetization \cite{tf},
  and we see how they evolve into 
fractional-charge vortices in the presence of rotation around $z$ direction given by the 
rotation operator \cite{ueda}
 \begin{align}
{\cal D}(\alpha)=\left( {\begin{array}{ccccc}
   e^{-2i\alpha} & 0&0&0&0\\0  &e^{-i\alpha}&0&0&0\\ 0&0&1&0&0\\
0&0&0&e^{i\alpha}&0 \\   0&0&    0 & 0&   e ^{2i\alpha} \      \end{array} } \right)
\end{align}
The first of these states has the wave-function components $\phi_{\pm 1}=\phi_0=0$,
and a general normalized wave function  under rotation can be obtained by operating 
${\cal D}(\alpha)$ on the representative state 
$\chi_0=(1/\sqrt 2,0,0,0,1/\sqrt 2)^T$
and
 has the form $e^{i\theta}(e^{-2i\alpha}/\sqrt 2,0,0,0, e^{2i\alpha}/\sqrt 2 )^T.$ The single-valuedness of the wave function with a fractional charge ($\theta \ne 0$) can be maintained if we take 
 $\theta=\beta/2, \alpha=\beta/4$ to get  $e^{i\beta/2}(e^{-i\beta/2}/\sqrt 2,0,0,0, e^{i\beta/2}/\sqrt 2 )  ^T,$ which is a 1/2-1/4 vortex, or a vortex of fractional charge 1/2.
 This is a vortex of charge {$0$} in component $\phi_{+2}$
and charge 1 in component $\phi_{-2}$.

The second degenerate   ground state of the  spin-2 polar BEC has 
$\phi_{\pm 2}=\phi_0=0$ and the  normalized wave function under rotation is
  $e^{i\theta}(0,e^{-i\alpha}/\sqrt 2,0, e^{i\alpha }/\sqrt 2,0 )  ^T$ \cite{tf}.  {This state 
is very similar to the spin-1 polar BEC state considered above.} 
This state will be single-{valued} for 
 $\theta =\alpha =\beta/2$ to get 
$e^{i\beta/2}(0,e^{-i\beta/2}/\sqrt 2,0, e^{i\beta/2}/\sqrt 2,0 ) ^T$ which is 
  a 1/2-1/2 vortex or a vortex of  charge 1/2. This generates a vortex of charge {$0$} in component $\phi_{+1}$
and charge 1 in component $\phi_{-1}$.

The third degenerate ground state of the  spin-2 polar BEC has $\phi_{\pm 2}= \phi_{\pm 1}=0$, and 
this possibility does not lead to a vortex of fractional charge.

(b) There are two degenerate ground states of a non-rotating spin-2 cyclic BEC \cite{tf},
and 
we consider the possibility of generating fractional-charge vortex from these states.  The first of these states has 
 wave-function components  $\phi_{ 1}=\phi_0=\phi_{-2}=0$, and a general 
normalized  wave function under rotation can be written as  
  $e^{i\theta}(e^{-2i\alpha}/\sqrt 3,0,0, e^{i\alpha}\sqrt{2/3},0 ) 
^T.$ There are {four} simple ways to maintain a single-valued spinor function to generate a fractional-charge vortex \cite{cyclic}. 
(i) One can  take $\theta=\beta/3,\alpha=2\beta/3  $,  to get   
$e^{i\beta/3}(e^{-4i\beta/3}\sqrt{1/3},0,0, e^{2i\beta/3}\sqrt{2/3},0 )^T$ 
which is a 1/3-2/3 vortex,  or a vortex of fractional charge 1/3.  
This is a vortex of  charge 1 in both $m_f =+2$ and $-1$ components.
(ii) The second choice  $\theta=-\beta/3,\alpha=\beta/3$ leads to   the wave function
$e^{-i\beta/3}(e^{-2i\beta/3}\sqrt{1/3},0,0, e^{i\beta/3}\sqrt{2/3},0 )^T$ 
which is  a 1/3-1/3 vortex, or a vortex of fractional charge 1/3. This is a vortex of charge 1 in component $\phi_{+2}$
and charge 0 in component $\phi_{-1}$. 
(iii) The third possibility is to take  $\theta=-2\beta/3, \alpha=2\beta/3, $  to obtain   
$e^{-i2\beta/3}(e^{-4i\beta/3}\sqrt{1/3},0,0, e^{2i\beta/3}\sqrt{2/3},0 )^T,$ 
which is a 2/3-2/3 vortex,  or a vortex of fractional charge 2/3.  This is a vortex of charge 2 in component $\phi_{+2}$
and charge 0 in component $\phi_{-1}$. 
(iv) The final choice  $\theta=2\beta/3,\alpha=\beta/3  $ yields the wave function   
$e^{i2\beta/3}(e^{-2i\beta/3}\sqrt{1/3},0,0, e^{i\beta/3}\sqrt{2/3},0 )^T$ 
which is a 2/3-1/3 vortex, or a vortex of fractional charge 2/3. This is a vortex of charge 0 in component $\phi_{+2}$ and charge 1  in component $\phi_{-1}.$


The other  possibility of the cyclic ground state has  $\phi_{\pm 1}=0$ and a general normalized rotating state has the form 
$e^{i\theta}(e^{-2i\alpha}/2,0,i/\sqrt 2,0, 
e^{i2\alpha}/2 )  ^T.$  However, this state does not lead to a vortex of fractional charge and will not be considered here.


 
In all the aforementioned cases of a vortex with a  fractional charge less than unity,  the spinor BEC has only two non-zero components.
In all these cases, the GP equations for the spinor BEC reduce
to that   for a binary BEC with components $j=1,2$ , e.g.,
\begin{align}\label{bi}
&i\frac{\partial}{\partial t}\phi_j =\left[-\frac{ {\cal O}^2}{2}+\frac{r^2}{2}
+ g_j\phi_j^2 + g_{12}\phi_{3-j}^2\right]\phi_j,\\
&{\cal O}^2 =\frac{1}{r}\frac{\partial}{\partial r} r 
\frac{\partial}{\partial r}.
\end{align} where $g_j$'s are the intra-component {nonlinearities} and 
$g_{12}$ is the inter-component nonlinearity.
{ For a  polar  spin-1 BEC  with a 1/2-1/2 vortex
(coupling $\phi_{\pm 1}$), these nonlinearities are defined as: 
\begin{equation} 
g_1 = g_2 = c_0 + c_1,~g_{12}=c_0-c_1.\label{nl4}
\end{equation}
Similarly, for a polar spin-2 BEC  hosting a
 1/2-1/4 vortex (coupling $\phi_{\pm 2}$)  or a
 1/2-1/2 vortex   (coupling $\phi_{\pm 1}$) 
{the nonlinearities
are}
\begin{align} 
g_1 =g_2&=c_0+4c_1,~g_{12}=c_0-4c_1+2c_2/5,\label{nl2}\\
g_1 =g_2&=c_0+c_1,~g_{12}=c_0-c_1+2c_2/5,\label{nl3}
\end{align}
{respectively.}
The same for a two-component spin-2 cyclic BEC (coupling $\phi_{+2}$ and $\phi_{-1}$) are
\begin{equation}
 g_1 = c_0 + 4c_1,~g_2 = c_0 + c_1,~ g_{12} = c_0 -2c_1. \label{nl1}
\end{equation}}
Using Eqs. (\ref{nl4})-(\ref{nl1}) as the definitions of intra- and 
inter-species interactions, the aforementioned spinor BECs with
 fractional-charge vortex can be treated as equivalent to a  binary BEC.

\section{Variational approximation  for the fractional-charge vortex}
\label{sec-IV}

In all the examples of fractional-charge vortex discussed in Sec. \ref{3}, 
we have two possibilities: (i)  a single- or doubly-charged vortex in one component of a 
binary BEC with components $j=1,2$,  and (ii) a singly-charged vortex in 
both components. First,   
let us assume a vortex  in one of the components only, 
say $j=2$. The binary GP equation for the components $\psi_j$ now has the form
\begin{align}\label{kt}
&i\frac{\partial}{\partial t}\psi_j =\left[-\frac{{\cal O} ^2}{2}+\frac{\delta_{j2}{\cal L}^2}
{2r^2}+\frac{r^2}{2}+ g_j|\psi_j|^2+ g_{12}|\psi_{3-j}|^2\right]\psi_j , 
\end{align}
where the Kronecker delta $\delta_{j2}$ sets the centrifugal term in the second component with angular momentum ${\cal L}=1,2$.
Equation (\ref{kt}) is applicable 
in all polar spin-1 and -2 BECs  described in Sec. III and  cyclic 1/3-1/3, 2/3-1/3 and 2/3-2/3 vortices. 
 We take 
 the variational wave functions as \cite{Perez-Garcia}
\begin{align}
&\psi_1=\frac{\sqrt{{\eta_1}}}{w_1\sqrt {\pi}} \exp\left[-\frac{r^2}{2w_1^2}+i\kappa_1 r^2\right]\\
&\psi_2=\frac{r^{\cal L}\sqrt{{\eta_2}}}{w_2^{{\cal L}+1}\sqrt{{\cal L} \pi}} \exp\left[-\frac{r^2}{2w_2^2}+i\kappa_2r^2\right]
\end{align}
with normalization $\int  |\psi_j|^2 d {\bf r} ={\eta_j}$.
Here $w_j$, the widths,  and $\kappa_j$, the chirps, are the variational parameters.  
The Lagrangian of the binary system is \cite{Perez-Garcia}
\begin{align}
L&= N \int \biggr[ \frac{1}{2} \biggr\{
 \frac{{\cal L}^2 |\psi_2|^2}{r^2} +\sum_j \biggr( i \big(\psi_j \frac{\partial}{\partial t} \psi_j^*-\psi_j^* \frac{\partial}{\partial t} \psi_j\big)
\nonumber \\ & + \biggr| \frac{\partial \psi_j}{\partial r}\biggr|^2
+ r^2|\psi_j|^2+ g_j|\psi_j|^4\biggr)\biggr\} 
 +  g_{12}   |\psi_1|^2|\psi_2|^2\biggr] d  {\bf r}\nonumber \\ 
&= N\Bigg[\frac{
\eta_1}{2}\biggr\{ w_1^2+\frac{1}{w_1^2}   \biggr\}  
+\frac{
\eta_2 ({\cal L}+1)}{2} \biggr\{ w_2^2+\frac{1}{w_2^2}   \biggr\}\nonumber \\
&
+\frac{
\eta_1^2g_1}{4\pi w_1^2}
+\frac{({\cal L}+1)
\eta_2^2g_2}{2^{{\cal L}+3} \pi w_2^2}+\frac{
\eta_1
\eta_2g_{12}w_1^{2{\cal L}}}{\pi(w_1^2+w_2^2)^{{\cal L}+1}}\nonumber \\ &
+
\eta_1w_1^2
(\dot \kappa_1+2\kappa_1^2)
+({\cal L}+1)
\eta_2w_2^2
(\dot \kappa_2+2\kappa_2^2)\Bigg].  
\end{align}
{The Euler-Lagrange equations for the variational parameters 
are
\begin{eqnarray}
\frac{d}{dt}\frac{\partial L}{\partial \dot w_j} -\frac{\partial L}{\partial w_j} &= 0,\\
\frac{d}{dt}\frac{\partial L}{\partial \dot \kappa_j} -\frac{\partial L}{\partial \kappa_j} &= 0,
\end{eqnarray}
which, for ${\cal L}=1$, can be simplified to
\begin{align}
2  (2 \kappa_1^2+\dot \kappa_1)+1 &= \frac{1}{w_1^4}+\frac{
\eta_1g_1}{2\pi w_1^4}+\frac{2
\eta_2g_{12}}{\pi}\frac{(w_1^2-w_2^2)}{(w_1^2+w_2^2)^3},\label{el2_a}\\
 2(2\kappa_2^2 + \dot \kappa_2)+1 &=\frac{1}{w_2^4}+\frac{
\eta_2g_2}{8\pi w_2^4}+
 \frac {2\eta_1g_{12}w_1^2}{\pi(w_1^2+w_2^2)^3},\label{el2_b}\\
\kappa_j &= \frac{\dot w_j}{2 w_j}\label{el2_c}.
\end{align}
From Eq. (\ref{el2_c}), $\dot \kappa_j$ is
\begin{equation}
\dot \kappa_j = -\frac{\dot w_j^2}{2 w_j^2}+\frac{\ddot w_j}{2 w_j}.\label{el2_cd}
\end{equation}
Using Eqs. (\ref{el2_c}) and (\ref{el2_cd}), we
eliminate $\kappa_j$ and $\dot\kappa_j$ from Eqs. (\ref{el2_a}) and (\ref{el2_b})
and obtain the following dynamical equations for the widths for ${\cal L}=1$:
\begin{align}
&\frac{\ddot w_1}{w_1}+1=\frac{1}{w_1^4}+\frac{
\eta_1g_1}{2\pi w_1^4}+\frac{2
\eta_2g_{12}}{\pi}\frac{(w_1^2-w_2^2)}{(w_1^2+w_2^2)^3},\label{el2a} \\
&\frac{\ddot w_2}{w_2}+1=\frac{1}{w_2^4}+\frac{
\eta_2g_2}{8\pi w_2^4}+\frac{2
\eta_1g_{12}}{\pi}\frac {w_1^2}{(w_1^2+w_2^2)^3}\label{el2b},
\end{align}
which are  to be solved numerically for studying the dynamics. For stationary results 
of densities, the second derivatives of widths are to be set to zero. 
Similarly, for ${\cal L}=2$, we have 
\begin{align}
&\frac{\ddot w_1}{w_1}+1=\frac{1}{w_1^4}+\frac{
\eta_1g_1}{2\pi w_1^4}+\frac{2
\eta_2g_{12}}{\pi}\frac{w_1^2(w_1^2-2w_2^2)}{(w_1^2+w_2^2)^4},\label{el2c} \\
&\frac{\ddot w_2}{w_2}+1=\frac{1}{w_2^4}+\frac{
\eta_2g_2}{16\pi w_2^4}+\frac{2
\eta_1g_{12}}{\pi}\frac {w_1^4}{(w_1^2+w_2^2)^4}\label{el2d}.
\end{align}

Next let us take  a vortex of unit charge in both components satisfying the binary GP equation
\begin{align}
&i\frac{\partial}{\partial t}\psi_j =\left[-\frac{{\cal O}^2}{2}+\frac{1}{2r^2}
+\frac{r^2}{2}+ g_j|\psi_j|^2+ g_{12}|\psi_{3-j}|^2\right]\psi_j. 
\end{align}
This is the case of a spin 2 cyclic 1/3-2/3 vortex.
In this case, we take the variational wave functions as 
\begin{align}
&\psi_j=\frac{r\sqrt{\eta_j}}{w_j^2\sqrt \pi} \exp\left[-\frac{r^2}{2w_j^2}+i\kappa_jr^2\right]. 
\end{align}
We recall that if $j=1$ represents the state $\psi_1=\phi_{+2}$ and $j=2$ represents the state  $\psi_2=\phi_{-1}$, then the condition of zero magnetization requires 
$\eta_2=2\eta_1=2/3$.
The Lagrangian is
\begin{align}
L&=N\int \biggr[ \frac{1}{2}
\sum_j \biggr\{ i (\psi_j \frac{\partial}{\partial t} 
\psi_j^*-c.c.)+\frac{1}{r^2}|\psi_j|^2 +  \left|\frac{\partial \psi_j}{\partial r}\right|^2\nonumber \\ 
&+  r^2|\psi_j|^2+ g_j|\psi_j|^4\biggr\}
  + g_{12}   |\psi_1|^2|\psi_2|^2\biggr] d\boldsymbol r \nonumber \\
&=\sum_j  N\biggr[ {\eta_j}\biggr( w_j^2+\frac{1}{w_j^2}   \biggr)    
+\frac{\eta_j^2g_j}{8\pi w_j^2}+2\eta_jw_j^2(\dot \kappa_j+2 \kappa_j^2)\biggr]\nonumber \\&
 +N\frac{2\eta_1\eta_2g_{12}w_1^2w_2^2}{\pi(w_1^2+w_2^2)^3}.
\end{align}
The Euler-Lagrange equations for the variational parameters lead to 
\begin{align}
&\frac{\ddot w_j}{w_j}+1=\frac{1}{w_j^4}+\frac{\eta_jg_j}{8\pi w_j^4}
+\frac{2\eta_{3-j}g_{12}}{\pi}\frac{w_{3-j}^2(2w_j^2-w_{3-j}^2)}{(w_1^2+w_2^2)^4}.
\label{el3}
\end{align}
If we recall Eq. (\ref{nl1}) and the zero magnetization condition $\eta_2=2\eta_1=2/3$ \cite{tf}, 
we find that $w=w_1=w_2$ is a solution of Eqs. (\ref{el3}), which reduces to 
\begin{align}\label{osc}
\ddot w +w=\frac{1}{w^3}\left(1+\frac{ c_0}{8\pi }\right)\equiv \frac{w_{\rm eq}^4}{w^3},
 \end{align}
where ${w_{\rm eq}}$ is the stationary width. For small oscillations $w(t)=w_{\rm eq}+X(t)$, where $X(t)$ is the deviation from the stationary value ${w_{\rm eq}}$.  We can linearize 
Eq. (\ref{osc}) in $X(t)$ to obtain 
\begin{equation}
\ddot X(t) = -\omega^2X(t), \quad \omega =2. \label{omega1}
\end{equation}
Hence, the components execute isotropic oscillation  
with a frequency which is twice of the trapping frequency.

}



\begin{figure}[t]
\begin{center}
\includegraphics[trim = 0mm 0mm 0cm 0mm, clip,width=0.85\linewidth,clip]{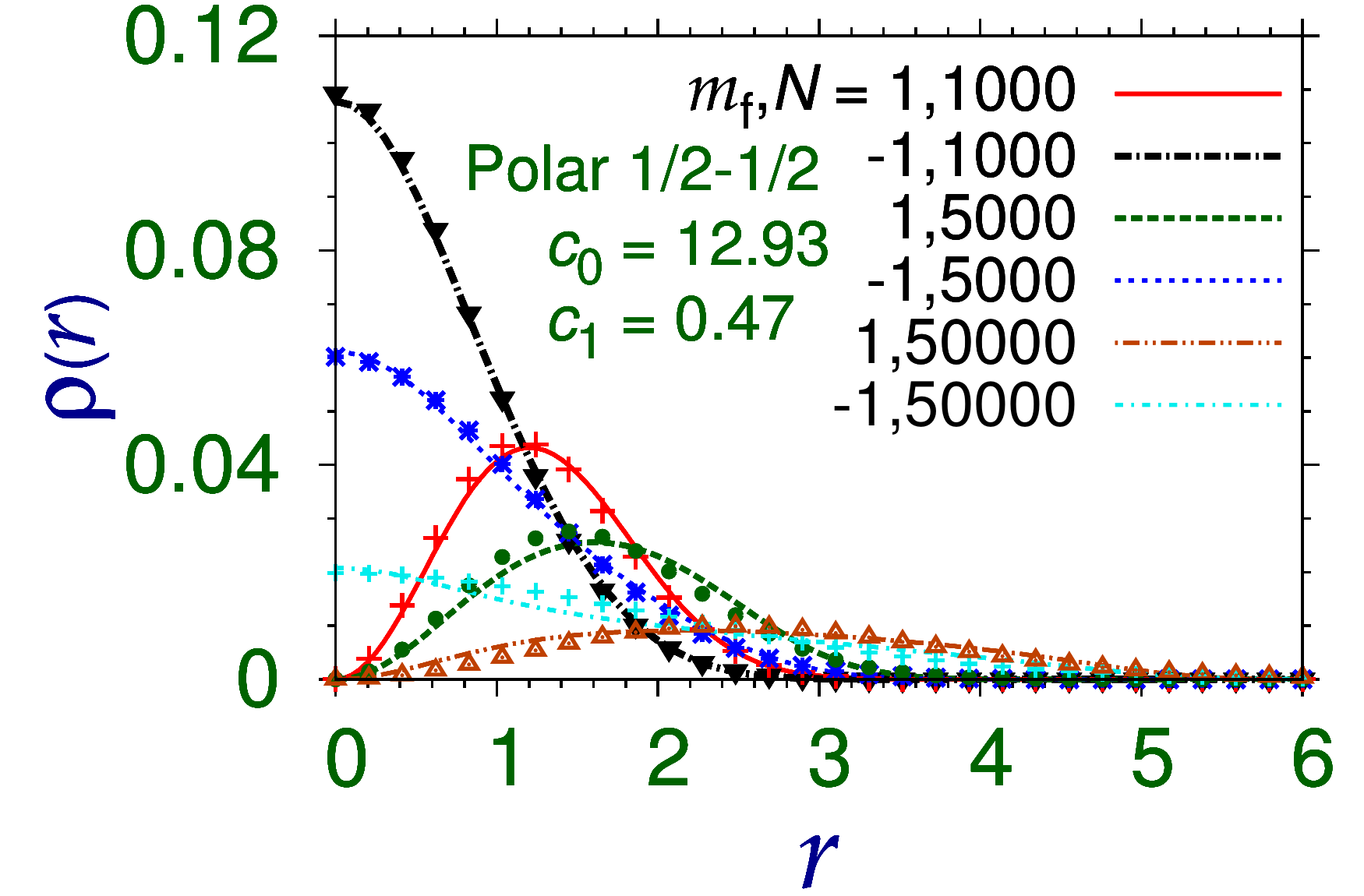}
 
\caption{(Color online)(a) Numerical (lines) and variational (points)   radial 
densities for 1/2-1/2 fractional-charge vortices of 1000, 5000, {and 50000} $^{23}$Na atoms  for  
components $\phi_{\pm 1}$ of a spin-1 polar BEC.  
 The fractional charge of the vortex is 1/2. The quoted values of $c_j$ are for 
$N=1000.$ The variables plotted in  this and other figures are dimensionless.
}
\label{fig-1} \end{center}
\end{figure}

\section{Numerical and variational results}
\label{V}

 We solve the coupled spin-1 and spin-2 GP equations 
numerically using split-step Crank-Nicolson scheme \cite{cn}. 
We employ the method suggested in Refs. \cite{num} for this purpose. 
We employ imaginary-time propagation for calculating stationary 
densities and real-time propagation for oscillation dynamics of the 
fractional-charge vortices.  We choose space and time steps of $0.05$ and
$0.00125$, respectively, for imaginary-time simulations, whereas for
the real-time simulations the respective values are $0.05$ and $0.0005$.
{ The confining trap frequencies are $\omega_x=\omega_y=2\pi\times20$ Hz and $\omega_z=2\pi\times400$ Hz.
We consider a spin-1 BEC of $^{23}$Na atoms, which has a polar ground state \cite{Stenger,ueda}, and
spin-2 BECs of $^{23}$Na   and $^{87}$Rb atoms, both of which have   polar ground
states \cite{ueda,Kuwamoto,Stamper-Kurn}, to investigate the fractional vortices. The units of
length for  $^{23}$Na and $^{87}$Rb spinor BECs considered in this paper are $l_0 = 4.69\mu$m and $l_0 = 2.41\mu$m,
respectively. In order to study vortices in the cyclic phase of spin-2 BEC, we
consider $^{23}$Na atoms with one of the scattering lengths suitably modified to
access the cyclic phase. 
}

\begin{figure}[t]
\begin{center}
\includegraphics[trim = 6mm 0mm 4mm 0mm, clip,width=0.49\linewidth,clip]{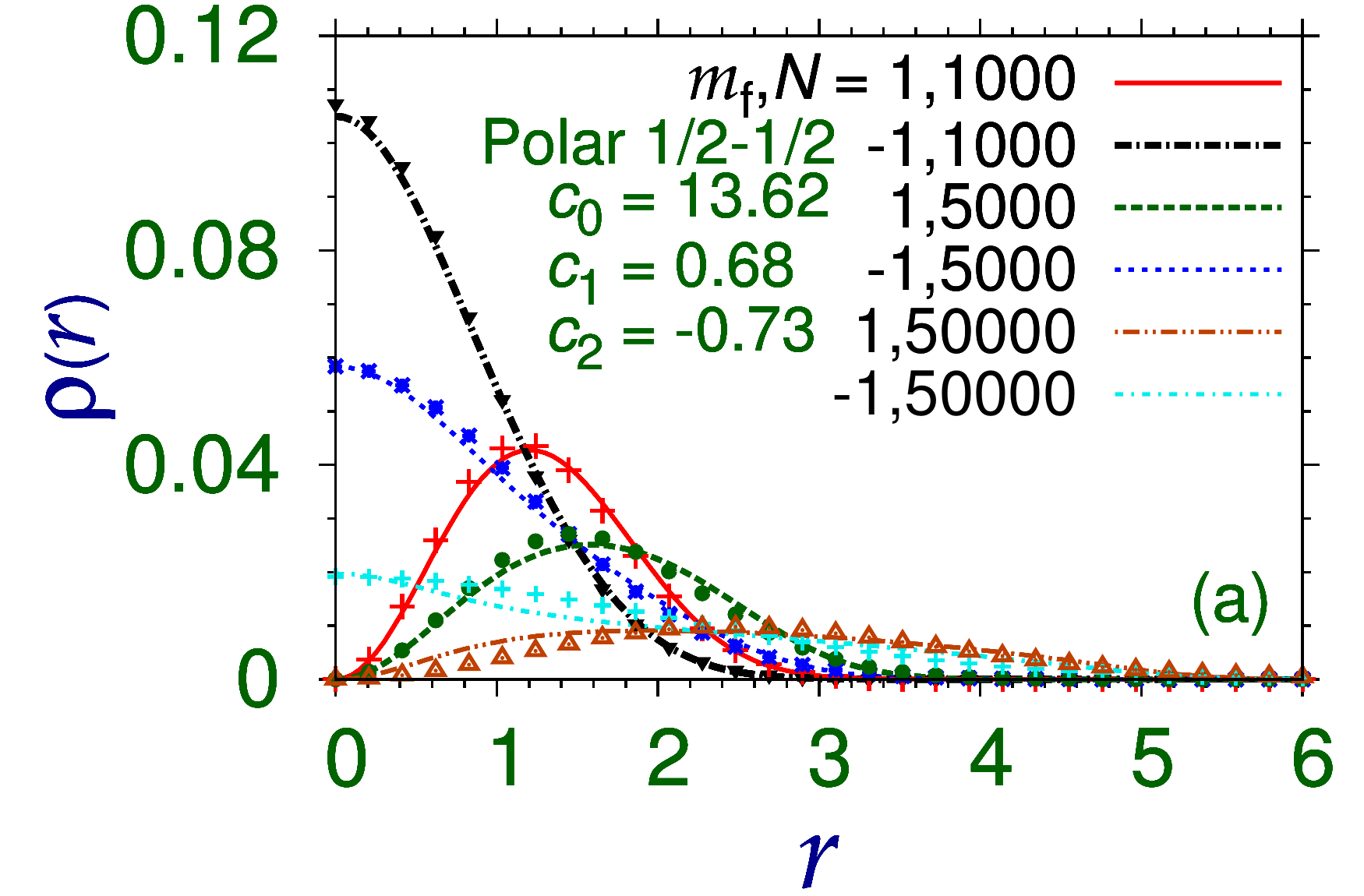}
\includegraphics[trim = 6mm 0mm 4mm 0mm, clip,width=0.49\linewidth,clip]{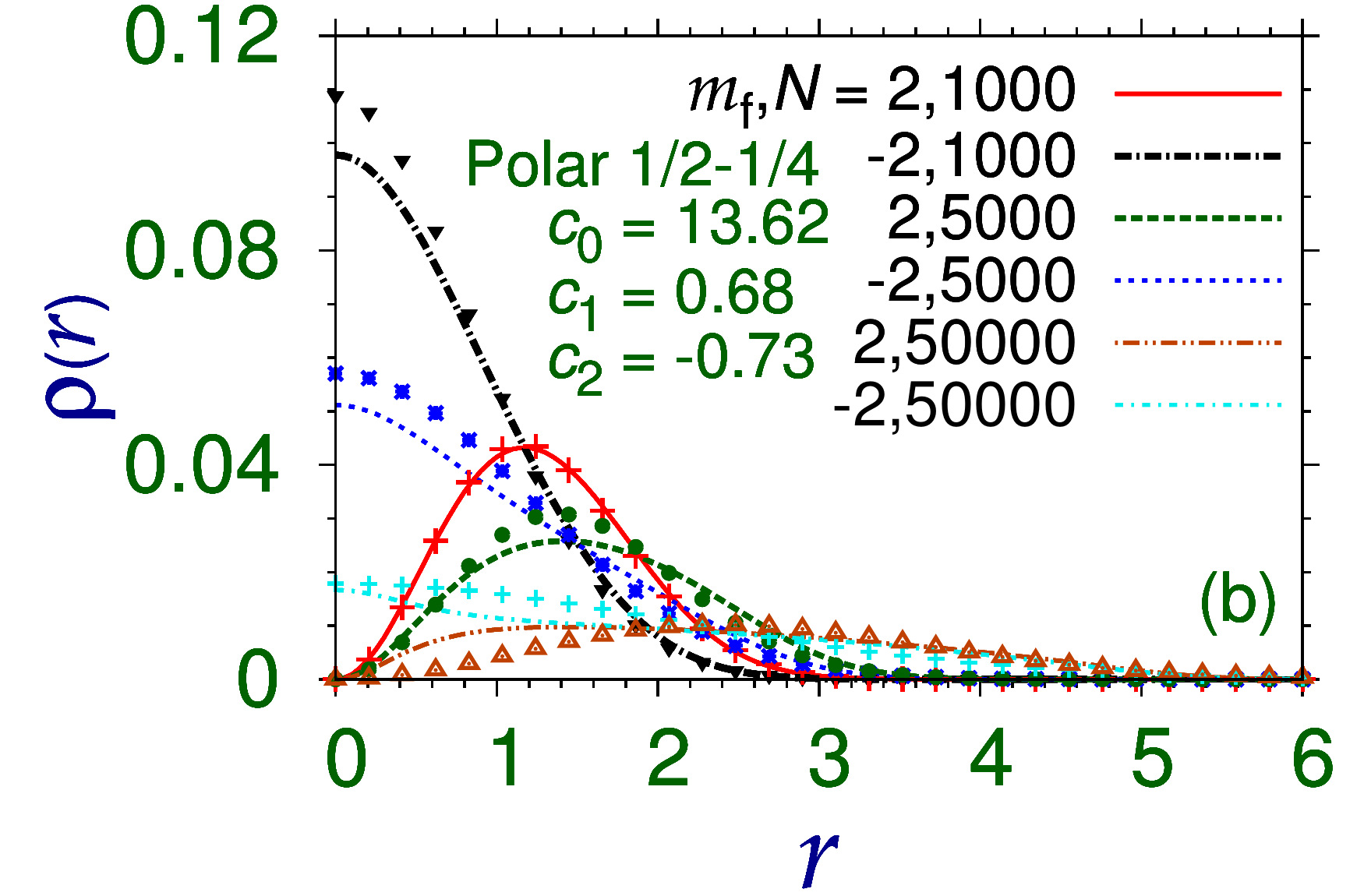}
\includegraphics[trim = 6mm 0mm 4mm 0mm, clip,width=0.49\linewidth,clip]{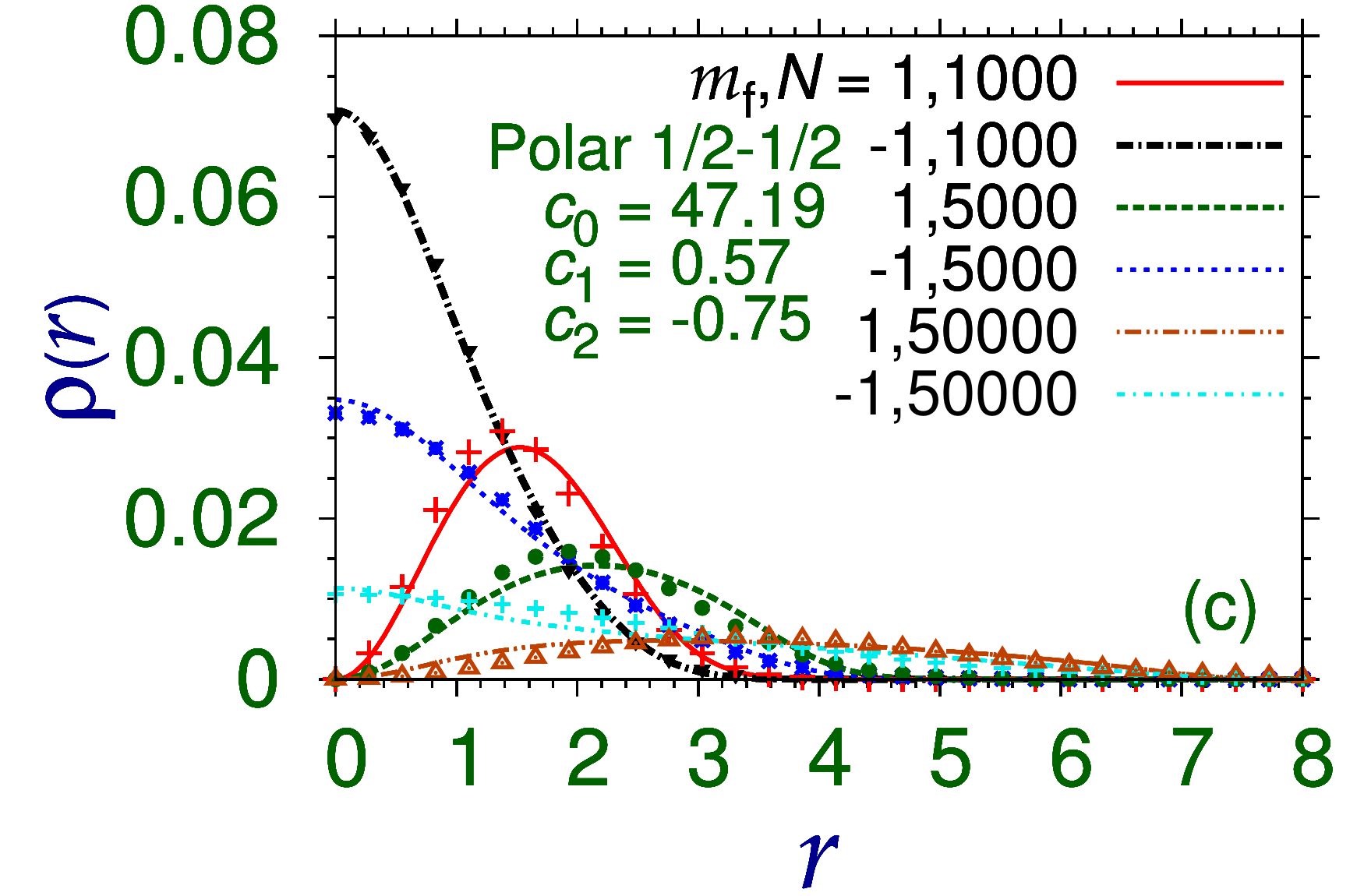}
\includegraphics[trim = 6mm 0mm 4mm 0mm, clip,width=0.49\linewidth,clip]{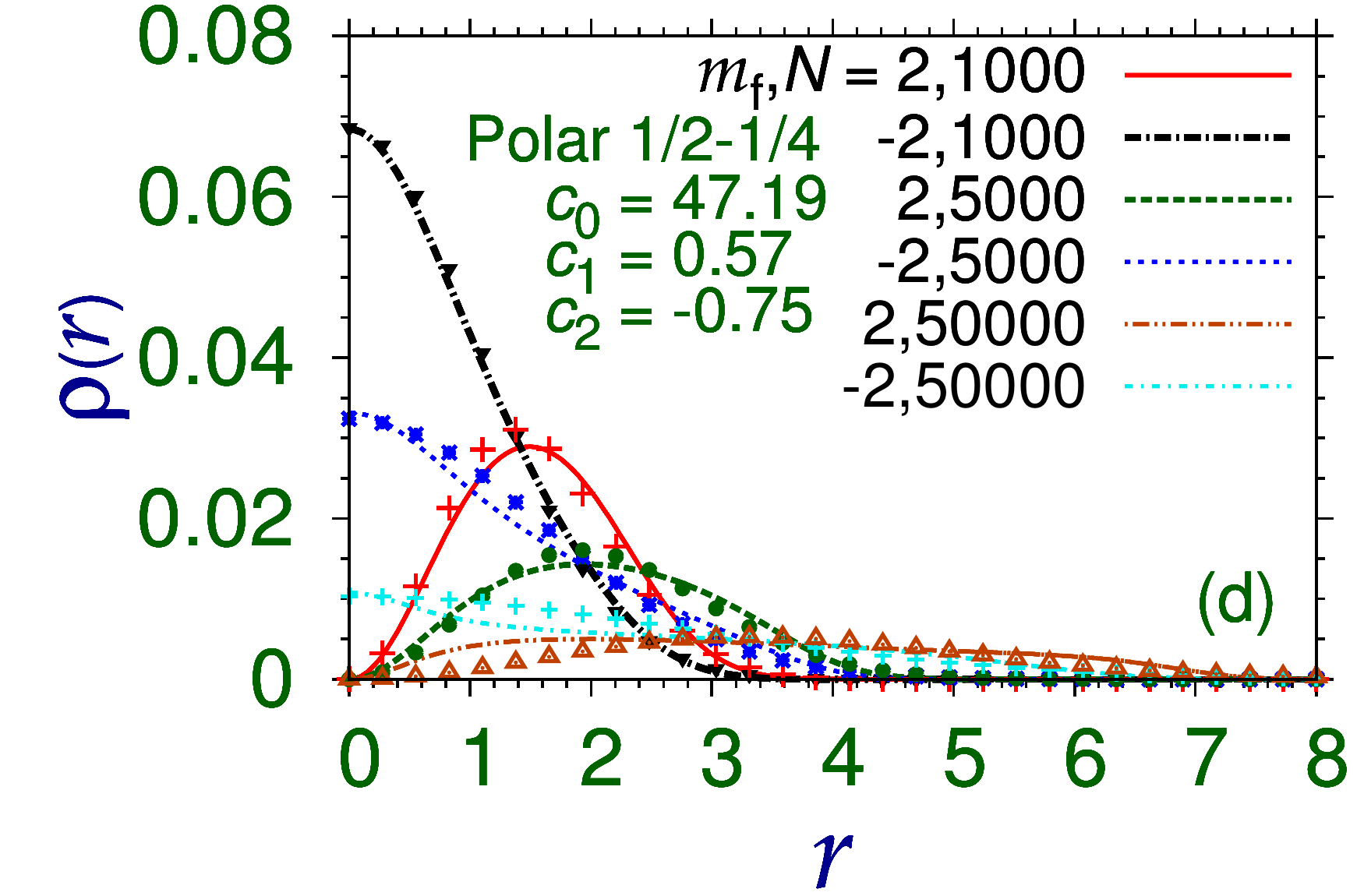}
 
\caption{(Color online)
Numerical (lines) and variational (points) radial densities 
for (a) 1/2-1/2  and  (b) 1/2-1/4 vortices in a polar spin-2 BEC of 1000, 5000, and 50000
$^{23}$Na atoms. The same for $^{87}$Rb atoms are shown in (c) and (d), respectively.
{
In the  case of 1/2-1/2 vortices
in (a) and (c), the densities correspond to components $\phi_{\pm 1}$ of spin-2 polar BECs, 
whereas in case of 1/2-1/4, these correspond to components $\phi_{\pm 2}$.} The fractional charge of all vortices is 1/2.  
The quoted values of $c_j$ are for 
$N=1000.$
}
\label{fig-2} \end{center}
\end{figure}

To study the fractional-charge vortex in a spin-1 polar system, we consider BECs of {1000, 5000 and $50000$} 
$^{23}$Na atoms with the experimental scattering lengths $a_0=47.36a_B$ and $a_2=52.98a_B$ \cite{ueda}, 
where $a_B$ is the Bohr radius. In these cases, {$l_0 = 4.69\mu$m and} the $(c_0,c_1)$ values are $(12.93,0.47)$, 
$(64.67,2.37)$, {and $(646.65,23.7)$,} respectively, for 1000, 5000, and {50000 atoms}.
The numerical and variational densities for the components $\phi_{\pm 1}$ are plotted in Fig. \ref{fig-1}. 
Here there is a singly-charged vortex in the component $\phi_{+1}$ and the variational results are calculated using 
Eqs. (\ref{el2a}) and (\ref{el2b}) setting the time derivatives equal to zero and $\eta_1=\eta_2 = 0.5$
in these equations. 

{Next, we consider spin-2 polar BECs of 1000, {5000, and 50000} 
$^{23}$Na and $^{87}$Rb atoms
with fractional-charge
vortices.} The three scattering lengths of $^{23}$Na are $a_0 = 34.9a_B$, $a_2 = 45.8a_B$,
and $a_4= 64.5a_B$ \cite{Ciobanu}{, whereas those for $^{87}$Rb are $a_0 = 87.4a_B$, $a_2 = 92.4a_B$,
and $a_4= 100.5a_B$ \cite{Stamper-Kurn}.} {Again, $l_0 = 4.69\mu$m for $^{23}$Na} and 
with $1000$, $5000$, and 50000 atoms the $(c_0,c_1,c_2)$ values
are $(13.62,0.68,-0.73)$, $(68.09,3.38,-3.65)$, {and $(680.9,33.8,-36.5)$,} respectively. 
{For $^{87}$Rb, $l_0 = 2.41\mu$m and
with $1000$, $5000$, and 50000 atoms the $(c_0,c_1,c_2)$ values
are $(47.19,0.57,-0.75)$, $(235.93,2.85,-3.76)$, {and $(2359.3,28.5,-37.6)$,} respectively.}
A polar spin-2 BEC with a fractional-charge vortex
is described  by the   binary GP equation (\ref{bi}) which is solved variationally using 
Eqs. (\ref{el2a}) and (\ref{el2b}) for this stationary problem.    
The numerical
and variational radial densities {in this case} for the 1/2-1/2, and 1/2-1/4 vortices in the polar spin-2 {BEC
of $^{23}$Na atoms}  
are shown in  \ref{fig-2}(a) and (b), respectively.  {The same for  polar spin-2 BEC
of $^{87}$Rb atoms are shown in Figs. \ref{fig-2}(c) and (d).} In the case {of 1/2-1/2 vortex}, 
there is a singly-charged vortex in the component $\phi_{+1},$ 
{ whereas} the vortex is in the component $\phi_{+2}$ {in the case of a 1/2-1/4 vortex}. 
In both the cases, the fractional charge of the vortex is 1/2.

\begin{figure}[t]
\begin{center}
\includegraphics[trim = 6mm 0mm 5mm 0mm, clip,width=0.49\linewidth,clip]{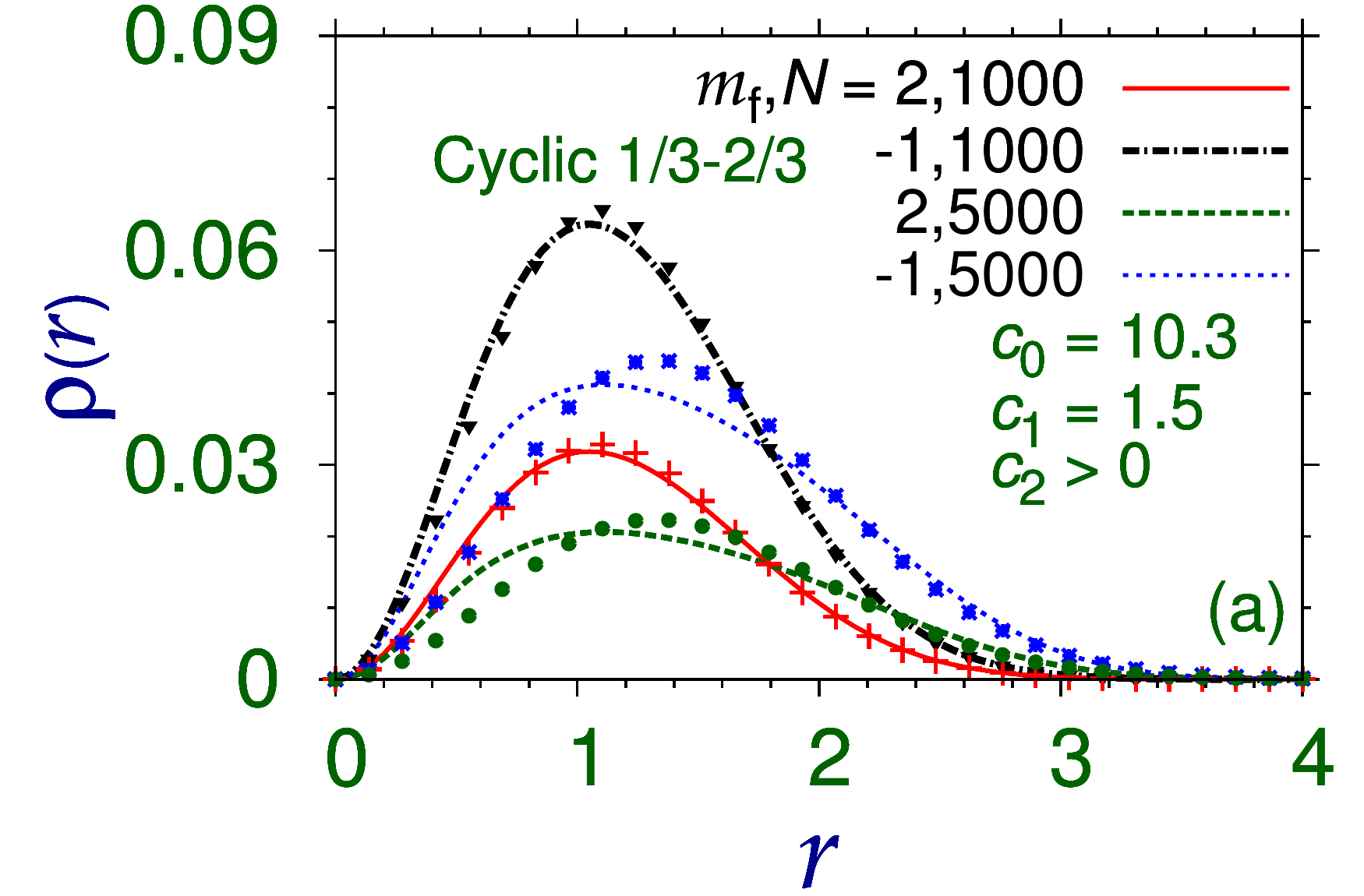}
\includegraphics[trim = 6mm 0mm 5mm 0mm, clip,width=0.49\linewidth,clip]{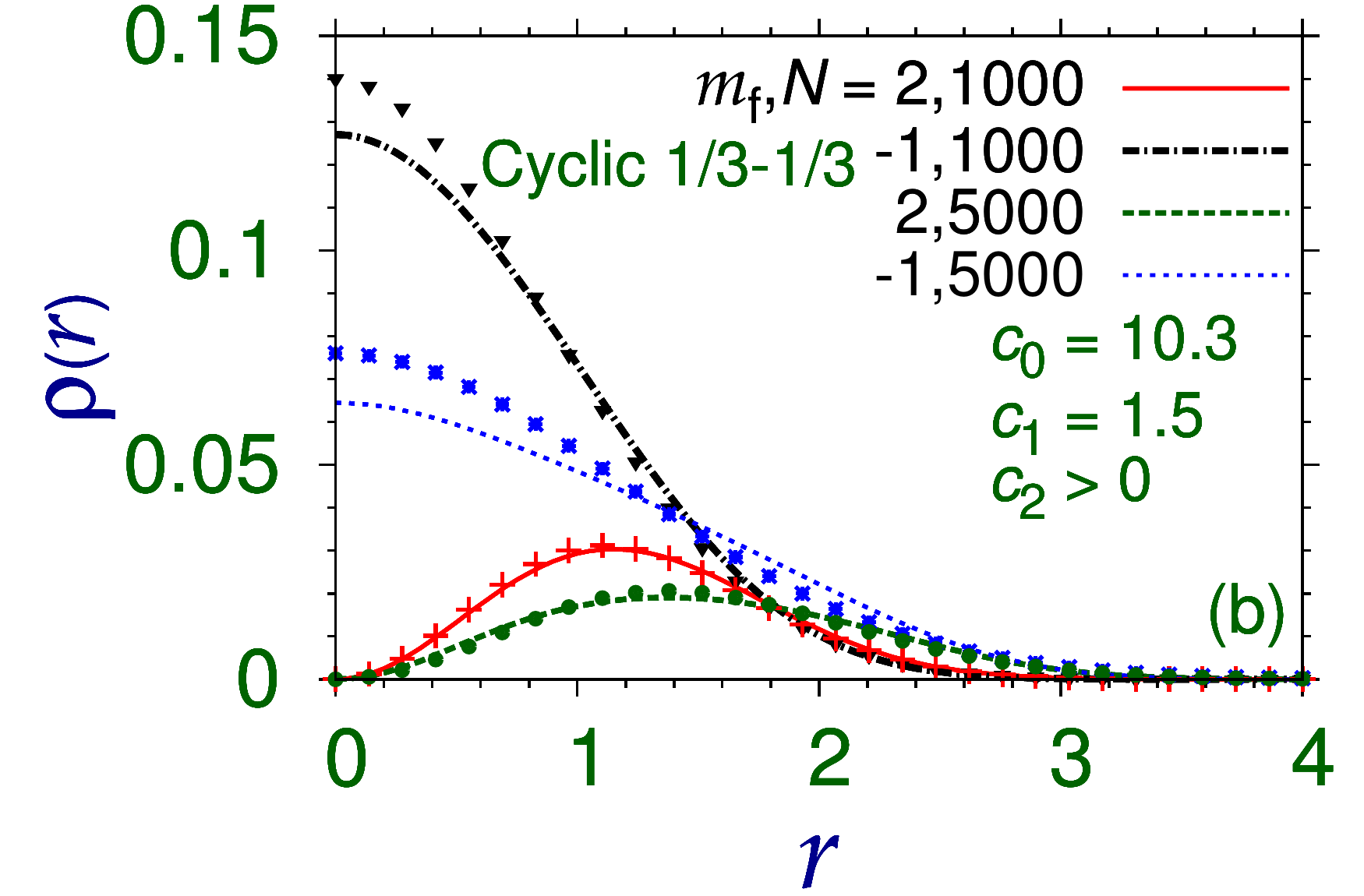}
\includegraphics[trim = 6mm 0mm 5mm 0mm, clip,width=0.49\linewidth,clip]{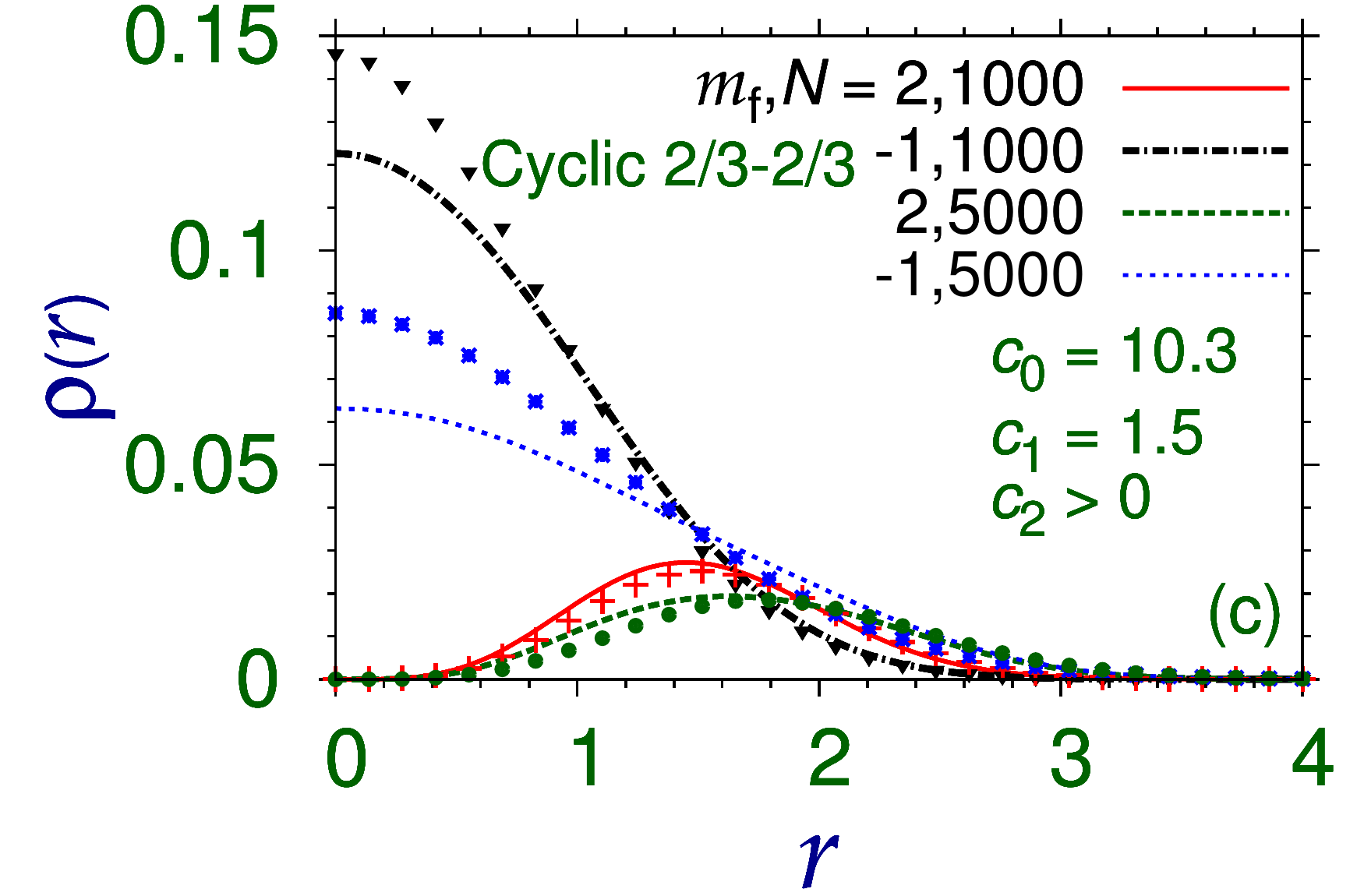}
\includegraphics[trim = 6mm 0mm 5mm 0mm, clip,width=0.49\linewidth,clip]{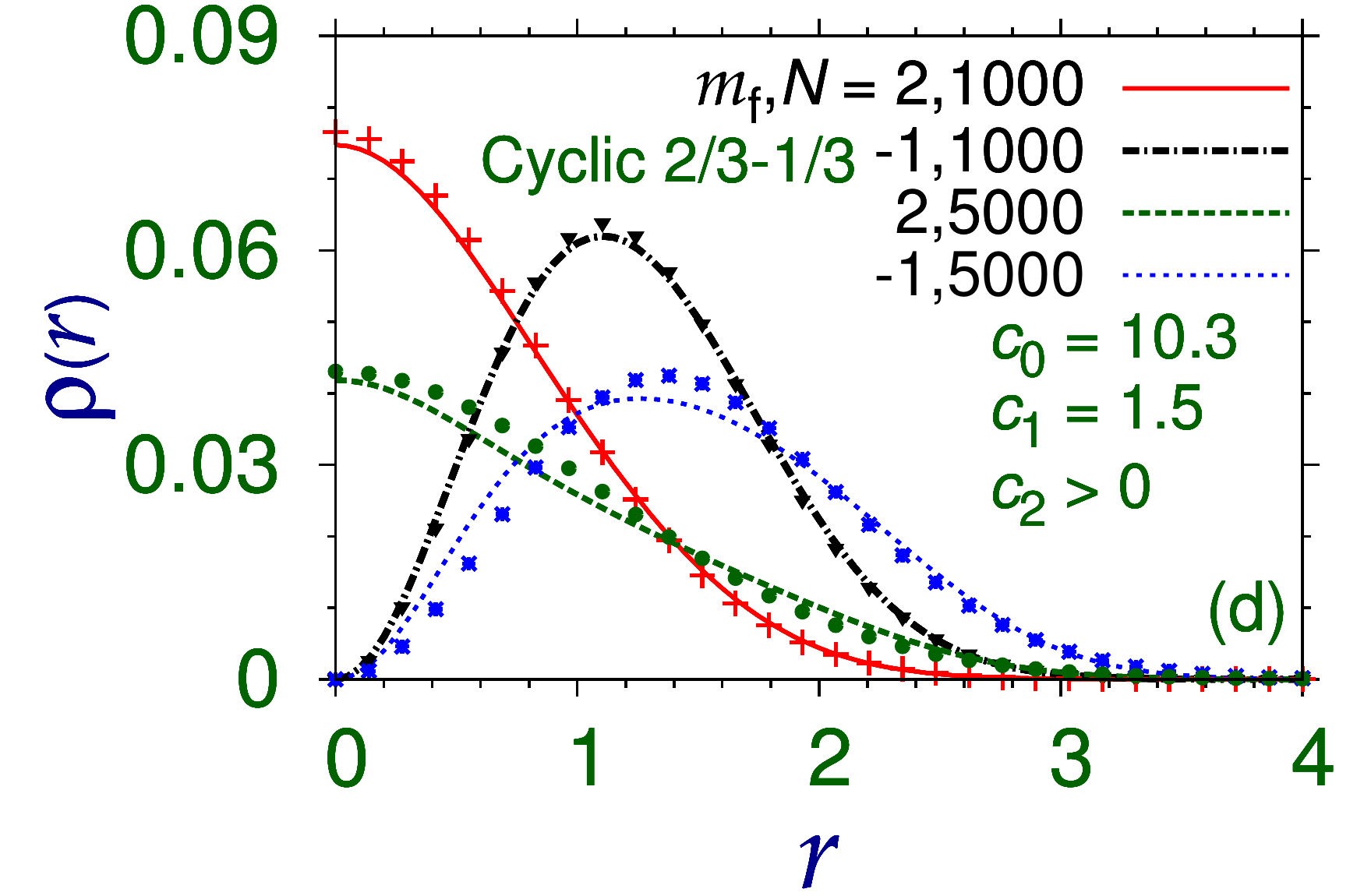}
\caption{(Color online) Numerical (lines) and variational (points)   radial 
densities of 1000 and 5000 $^{23}$Na atoms  corresponding to  
components $\phi_{+ 2}$ and $\phi_{-1}$ of a cyclic spin-2 BEC generating (a) a 1/3-2/3 vortex, 
(b) a 1/3-1/3 vortex, (c) a 2/3-2/3 vortex, and (d) a 2/3-1/3 vortex.
In  (a) and (b)  the fractional charge of the vortices is 1/3 and in (c) and (d) it is 2/3. The quoted values of $c_j$ refer to $N=1000.$
}
\label{fig-3} \end{center}
\end{figure}

For a spin-2  cyclic system  with a fractional-charge vortex, we 
  consider a BEC of  $1000$ and $5000$ $^{23}$Na atoms with $a_0 = 34.9a_B$, 
$a_2 = 22.9a_B$, and $a_4= 64.5a_B$, where $a_2$ has been modified from its 
experimental value of $45.8a_B$ to access the cyclic phase. This can be achieved experimentally 
by exploiting
Feshbach resonances \cite{Inouye}. The numerical and variational  densities  
  for a 1/3-2/3, 1/3-1/3,  2/3-2/3, and 2/3-1/3   vortex of a   cyclic spin-2 BEC are
shown in Figs. \ref{fig-3}(a), (b), (c), and (d), respectively. 
The variational approximation  for the 1/3-2/3 vortex in Fig. \ref{fig-3}(a) was 
obtained using Eq. (\ref{el3}), {the same    for the 1/3-1/3, 2/3-1/3,    vortices  
in Figs. \ref{fig-3} (b) {and} (d)  were obtained using  Eqs. (\ref{el2a})-(\ref{el2b}), and
for the 2/3-2/3 vortex was obtained using Eqs. (\ref{el2c})-(\ref{el2d}).}  
The fractional charge of the vortices in Figs.  \ref{fig-3}(a) and  (b) is 1/3 and that in 
Figs.  \ref{fig-3}(c) and  (d) is 2/3. All the component vortices in Fig. \ref{fig-3} have charge one except the one in \ref{fig-3}(c) which has charge $2$. 

\begin{figure}[!t]
\begin{center}
\includegraphics[trim = 15mm 10mm 5.cm 15mm, clip,width=.9\linewidth,clip]{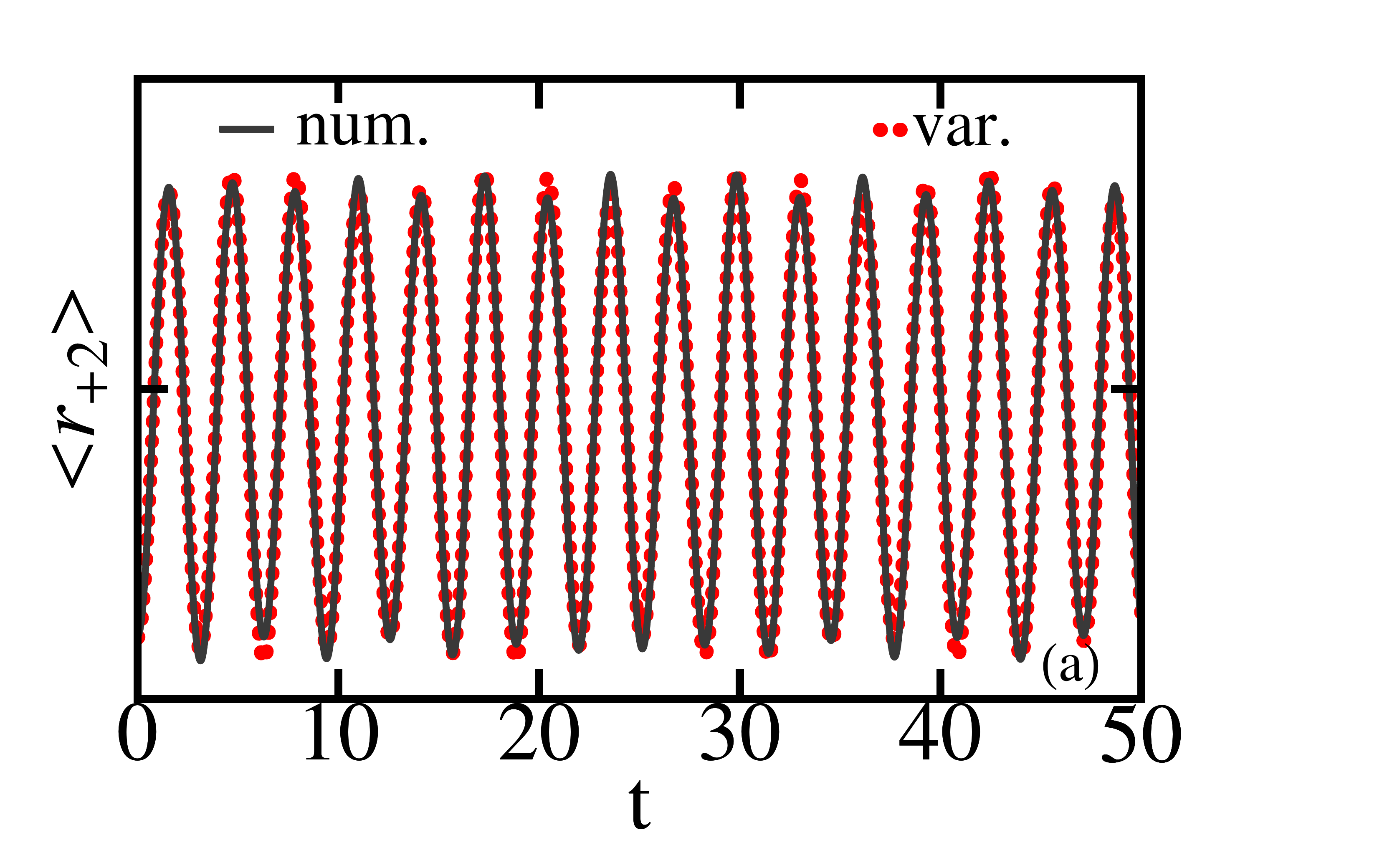}
\includegraphics[trim = 15mm 10mm 5.cm 15mm, clip,width=.9\linewidth,clip]{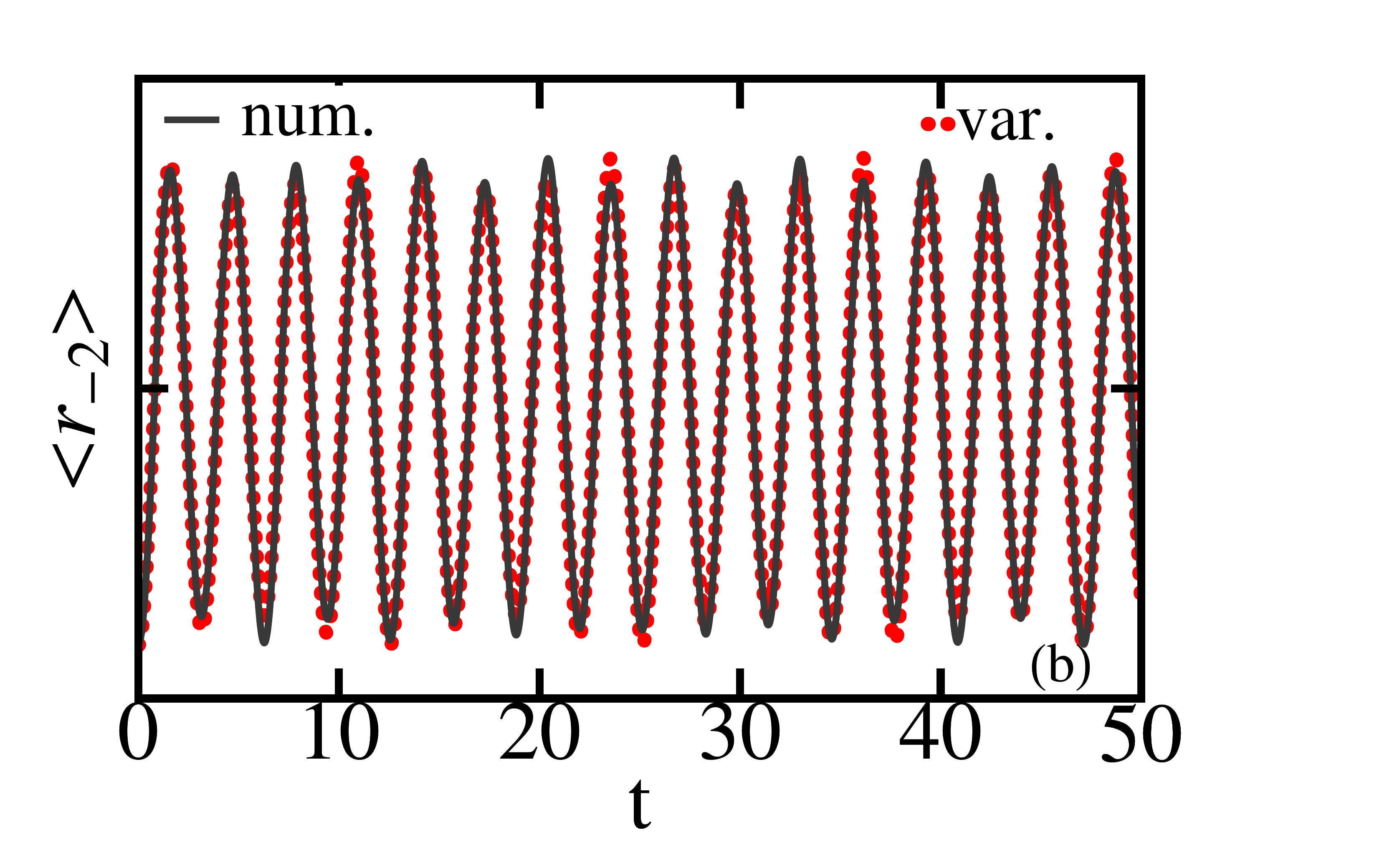}
\caption{(Color online) Numerical (lines) and variational (points) results
of 
collective-mode oscillation  of the rms sizes $\langle r \rangle$ of components (a) $m_f=+ 2$
and (b) $m_f=-2$
{in a spin-2 polar $^{23}$Na BEC of 50000 atoms hosting a 1/2-1/4 vortex. Numerically, the mode was excited}
by applying the perturbed potential 
(\ref{pert}). The stationary density of this  vortex is shown in Fig. \ref{fig-2}(b).  
The equivalent binary system has a {singly-charged} vortex in one of the components, $m_f=+2$.
}
\label{fig-4} \end{center}
\end{figure}

\begin{figure}[!t]
\begin{center}
\includegraphics[trim = 0mm 10mm 0cm 10mm, clip,width=\linewidth,clip]{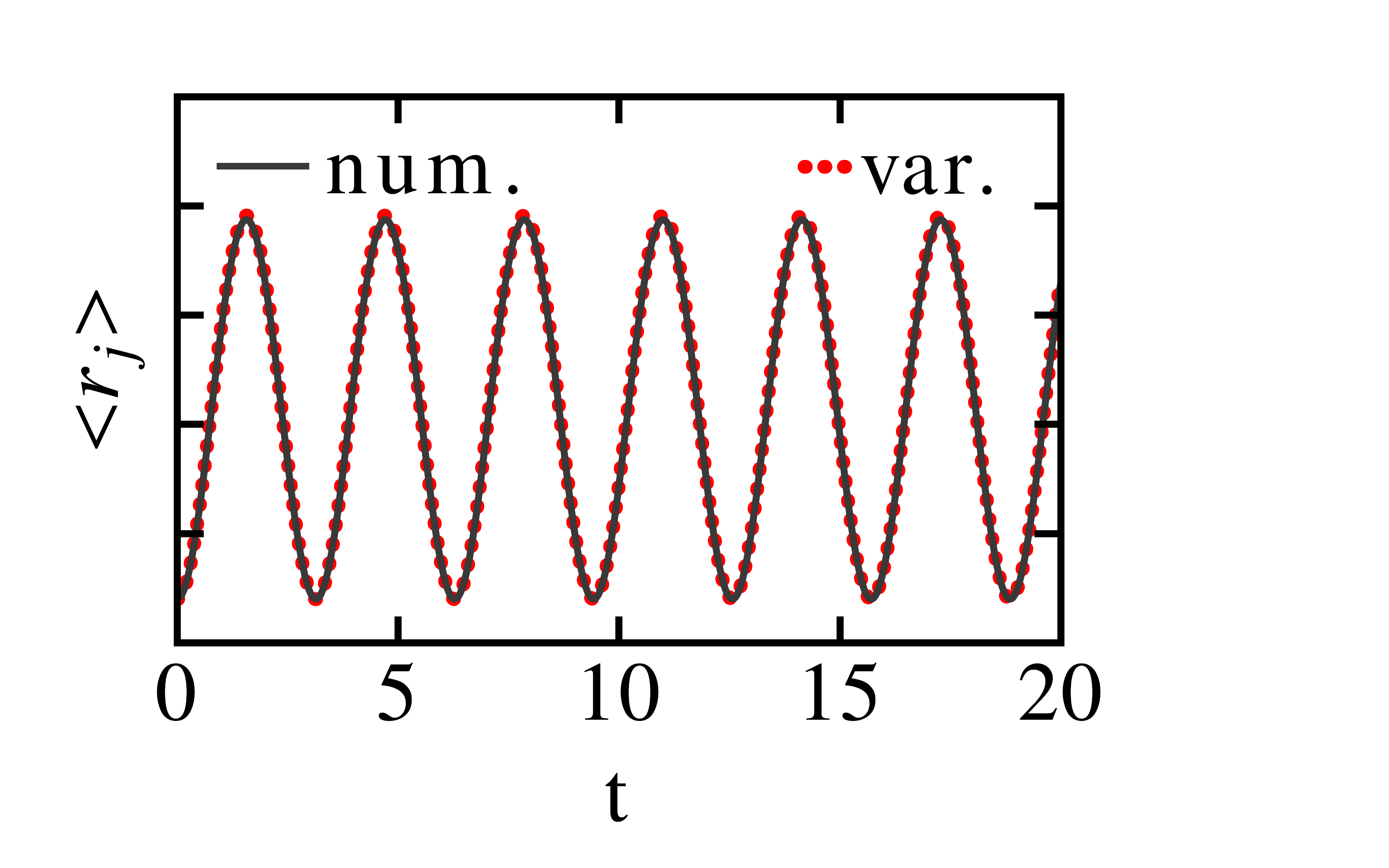}
\caption{(Color online) Numerical (lines) and variational (points) results
of 
collective-mode oscillation  of the rms sizes $\langle r \rangle$ of components $m_f=+2,-1$
{in a spin-2 cyclic
$^{23}$Na BEC of 50000 atoms hosting a 1/3-2/3 vortex. Numerically, the mode was excited} 
by  multiplying the nonlinearities $c_j$ by the constant factor {1.095} in real-time simulation. 
The stationary density of this  vortex is shown in Fig. \ref{fig-3}(a).  
The equivalent binary system has a singly-charged  vortex in both components, $m_f=+2, -1$.
}
\label{fig-5} \end{center}
\end{figure}

The low-energy collective modes  can be excited in a spinor BEC
by introducing a
small time-dependent sinusoidal perturbation to potential $V$ during some finite time
interval \cite{Jin,Dalfovo}.
In order to excite in-phase isotropic oscillations in a binary
system, which corresponds to the two components executing breathing mode
oscillations in phase, we apply the trapping potential 
\begin{equation}\label{pert}
V = \frac{1}{2}[x^2\{1 + \epsilon \cos(\omega_d t)\}+y^2\{1 + \epsilon \cos(\omega_d t)\}]
\end{equation}
 at time $t=0$
in  real-time simulation, where the strength of the perturbation $\epsilon\ll 1$.
We consider the driving frequency $\omega_d = \sqrt{2}$, and the perturbation is switched off  
($\epsilon=0$) later at  $t = 20$.
The numerical data for the collective oscillations is sampled after this period.
This mode can be easily excited in the experiments on the spinor BECs with
fractional-charge vortices where the trapping potential is the same for the two non-zero
components. An alternative method to excite this mode could be to increase or decrease
each of the nonlinearities $c_i$ in the GP equations by the same factor. To compare the numerical and variational
results for dynamics, we consider two examples: (a) 1/2-1/4  vortex in polar spin-2
  $^{23}$Na  BEC and (b) 1/3-2/3  vortex in the cyclic spin-2 
$^{23}$Na  BEC. The number of atoms considered in each case is {50000}.
In case (a), the collective mode is excited by modifying the trapping potential according
to the aforementioned prescription with {$\epsilon = 0.001$} in real-time simulation. The numerical and variational
results for the root mean square (rms) sizes
of the components are shown in Figs. \ref{fig-4}(a) and (b) for components 
$m_f=+2$ and $-2$, respectively.
The variational results correspond to the solutions of the coupled Eqs. (\ref{el2a})-(\ref{el2b}).
The numerical data for the
oscillation  is sampled from $t = 25$ onwards to remove a transient period after the perturbation
is switched off. The variables on the ordinate ($y$ axis) in Figs. \ref{fig-4}(a) and (b) are not shown explicitly as a shift of the variational results has been given 
in this direction before plotting. It is noteworthy that the variational and numerical frequencies 
coincide.

In case (b), the collective mode is excited by increasing all $c_i$'s slightly by the same factor {1.095}   at $t=0$ in 
real-time simulation.  The variational and numerical results for the rms sizes 
in this case
are shown in Fig. \ref{fig-5}. It has been shown \cite{tf} that the densities 
for the $m_f=+2$ and $m_f=-1$ states become multiple of each other ($\rho_{-1}=2\rho_{+2}$), 
and the GP equation is completely determined by the constant $c_0$ and is 
independent of $c_1$ and $c_2$ as in the single-mode approximation \cite{sma}.
Consequently, 
the rms sizes for the two components become identical, and they execute in-phase oscillation. 
Hence a single size is shown in Fig. \ref{fig-5}. 
The variational dynamics in this case corresponds to the solution
of Eqs. (\ref{el3}). {The numerical value of oscillation frequency $2.0$ is in excellent agreement
with the analytic result ($\omega=2$) in Eq. (\ref{omega1})}.  The same dynamics can also be observed by adding a time-dependent
perturbation to the trapping potential for a finite interval of time in real-time simulation.

\section{Summary and discussion}
\label{VI}

A scalar BEC can only have a vortex of integer angular momentum or charge.
On the other hand, a spinor BEC can have a fractional-charge vortex. The simplest of these 
has fractional charge less than unity. We classify all possible fractional-charge vortices of 
charge less than unity in spin-1 and spin-2 BECs. These vortices involve only two non-zero 
spin components of the spinor BEC, between which one or both components may {exhibit} a vortex.  
The statics and dynamics of these  vortices are studied employing an accurate numerical solution and a Gaussian variational approximation to a mean-field GP equation. The numerical and variational results for  stationary densities and frequencies of some collective-mode oscillation of these fractional-charge vortices are in good agreement with each other. These fractional charge-vortices can be studied experimentally and the predictions of the present study verified.

The present study also opens up several future directions of research. 
A spin-3 BEC has a much richer phase diagram {possibly} involving new phases and will allow different types of fractional-charge vortices with different charge than reported here.  The evolution of the present fractional-charge vortex in the presence of spin-orbit coupling and/or Rabi coupling would be an interesting topic of future investigation. Also,  fractional-charge vortices of charge larger than unity in spin-1 and spin-2 BEC might lead to some interesting features 
not seen before. Also, the dynamical stability of these vortices of charge larger than unity 
would be an interesting topic of future investigation.

\begin{acknowledgments}
{This investigation is supported in part by the FAPESP (Projects: 2013/07213-0 and  	
2012/00451-0) and CNPq (Project: 303280/2014-0) (Brazil). }
\end{acknowledgments}

\end{document}